\documentclass[twoside,a4paper]{article}
\usepackage{float}
\usepackage{graphicx}
\usepackage{fancyhdr}
\usepackage{multicol}
\usepackage{styl-ap}
\usepackage{ulem}

\begin{document}

\title{Clustering Measurements of broad-line AGNs: Review and Future}

\author{Mirko Krumpe\work{1,3}, Takamitsu Miyaji\work{2,3}, Alison L. Coil\work{3}}
\workplace{European Southern Observatory, 
Karl-Schwarzschild-Stra\ss e 2, 85748 Garching bei M\"unchen, Germany
\next
Instituto de Astronom\'a, UNAM, Apdo. Postal 106, Ensenada, BC, M\'exico
\next
University of California at San Diego, CASS, 9500 Gilman Drive, La Jolla, CA
92093-0424, USA}
\mainauthor{mkrumpe@eso.org}
\maketitle

\begin{abstract}%
Despite substantial effort, the precise physical processes that lead to 
the growth of super-massive black holes in the centers of galaxies are still 
not well understood. These phases of black hole growth are thought to be of key 
importance in understanding galaxy evolution. Forthcoming missions such as 
eROSITA, HETDEX, eBOSS, BigBOSS, LSST, and Pan-STARRS 
will compile by far the largest ever Active Galactic Nuclei (AGNs) catalogs 
which will allow us to measure the spatial distribution of AGNs in the 
universe with 
unprecedented accuracy. For the first time, AGN clustering measurements will 
reach a level of precision that will not only allow for an alternative approach 
to answering open questions in AGN/galaxy co-evolution but will open a new 
frontier, allowing us to precisely determine cosmological parameters. 
This paper reviews the large-scale clustering measurements of broad line AGNs. 
We summarize how clustering is measured and which constraints 
can be derived from AGN clustering measurements, we discuss recent 
developments, and we briefly describe future projects that will deliver 
extremely large AGN samples which will enable AGN clustering measurements 
of unprecedented accuracy. In order to maximize the scientific return on the 
research fields of AGN/galaxy evolution 
and cosmology, we advise that the community develop a full understanding 
of the systematic uncertainties which will, in contrast to today's measurement, 
be the dominant source of uncertainty. 

\end{abstract}


\begin{multicols}{2}
\section{Introduction}
Large area surveys such as the Two Degree Field Galaxy Redshift Survey  
(2dFGRS; Colless et~al.~2001) and Sloan Digital Sky Survey (SDSS;
Abazajian et~al.~2009) 
have measured positions and redshifts of millions of galaxies.
These measurements allow us to map the 3D structure 
of the nearby universe\footnote{A visual impression is given in this video: 
http://vimeo.com/4169279}. 

Galaxies are not randomly distributed in space. They form a complex 
cosmic network of galaxy clusters, groups, filaments, isolated 
field galaxies, and voids, which are large regions of space that
are almost devoid of galaxies. The current understanding of the distribution of 
galaxies and 
structure formation in the universe is based on the theory of gravitational
instability. 
Very early density fluctuations became the ``seeds'' of cosmic
structure. These have been 
observed as small temperature fluctuations 
($\delta T/T \sim 5\times 10^{-5}$) in the 
cosmic microwave background with the Cosmic 
Background Explorer (Smoot et~al.~1992). 
The small primordial matter density enhancements have progressively 
grown through gravitational collapse and created the complex network 
seen in the distribution of matter in the later universe. 

During a galaxy's lifetime different physical processes, which are still 
not well 
understood, can trigger a mass flow onto the central super-massive black hole 
(SMBH). 
In this phase of galaxy evolution, the galaxy is observed as an Active Galactic
Nucleus (AGN). 
After several million years, when the SMBH has consumed its accretion reservoir, the
central 
engine shuts down, and the object is again observed as a normal galaxy. The AGN
phase 
is thought to be a repeating special epoch in the process of galaxy 
evolution. 
In recent years it has become evident that both fundamental galaxy and AGN parameters 
change significantly between 
low ($z < 0.3$) and intermediate redshifts ($z\sim 1-2$), e.g., global star
formation density (Hopkins \& Beacom~2006) 
and accretion rate onto SMBHs. For example, the contribution
to black hole growth has shifted from high luminosity objects at high 
redshifts to low luminosity objects at low redshifts 
(AGN ``downsizing''; e.g., Hasinger et~al.~2005). 
It has also become clear that SMBH masses follow a 
tight relation with the mass or velocity dispersion of the stars in 
galactic bulges (Magorrian et~al.~1998; Gebhardt et~al.~2000;
Ferrarese \& Merritt~2000).
These observational correlations motivate a co-evolution scenario for 
galaxies and AGNs and provide evidence of a possible interaction or 
feedback mechanism between the SMBH and the host galaxy. 
The interpretation of this correlation, i.e., whether and to what
extent the 
AGN influences its host galaxy, is controversially debated (e.g., Jahnke \& Macci\'o~2011). 

Since AGNs are generally much brighter than (inactive) galaxies, one major
advantage of
AGN large-scale (i.e., larger than the size of a galaxy) clustering measurements over galaxy clustering measurements 
is that they allow the study of the matter distribution in the universe out to higher 
redshifts. At these very high redshifts, it becomes challenging and observationally 
expensive to detect galaxies in sufficient numbers. Furthermore, as the distribution 
of AGNs and galaxies in the universe depends on galaxy evolution physics,
large-scale 
clustering measurements are an independent method to identify and constrain the
physical 
processes that turn an inactive galaxy into an AGN and are responsible for  
AGN/galaxy co-evolution. 

In the last decade the scientific interest in AGN large-scale 
clustering measurements has increased significantly. As only a very small
fraction of 
galaxies contain an AGN ($\sim$1\%), the remaining and dominating 
challenge in deriving 
physical constraints based on AGN clustering measurements is the 
relative small sample size compared to galaxy clustering measurements. However, this 
situation will change entirely in the next decade when several
different 
surveys come online that are expected to identify millions of AGN 
over $\sim$80\% of cosmic time.

We therefore review the current broad-line AGN clustering measurements. 
A general introduction to clustering measurements is given in Sections~2 \& 3. In Section~4 
we briefly summarize how AGN clustering measurements have evolved and discuss 
recent developments. In Section~5, we discuss the outlook for 
AGN clustering measurements in future upcoming projects.

\section{Understanding Observed Clustering Properties}

In our current understanding, the observed galaxy and AGN spatial 
distribution in the universe -- i.e., large-scale clustering -- 
is caused by the interplay between cosmology and the physics of 
AGN/galaxy formation and evolution.

In the commonly assumed standard cosmological model, Lambda-CDM, the
universe is currently composed 
of $\sim$70\% dark energy, $\sim$25\% dark matter (DM),
and $\sim$5\% baryonic matter (Larsen et~al.~2011). Dark 
matter plays a key role in structure formation as it is the dominant 
form of matter in the universe.
Baryonic matter settles in the deep gravitational potentials created by dark
matter, the so-called dark matter halos (DMHs). 
The term ``halo'' commonly refers to a bound, gravitationally collapsed 
dark matter structure which is approximately in dynamical equilibrium. 
The parameters of the cosmological model 
determine how the DMHs are distributed in space (Fig.~\ref{fig1}, 
left panel, A-branch) as a function of the DMH mass and cosmic time. Different
cosmological 
models lead to different properties of the DMH population.

\begin{myfigure}
\centerline{\resizebox{83mm}{!}{\includegraphics{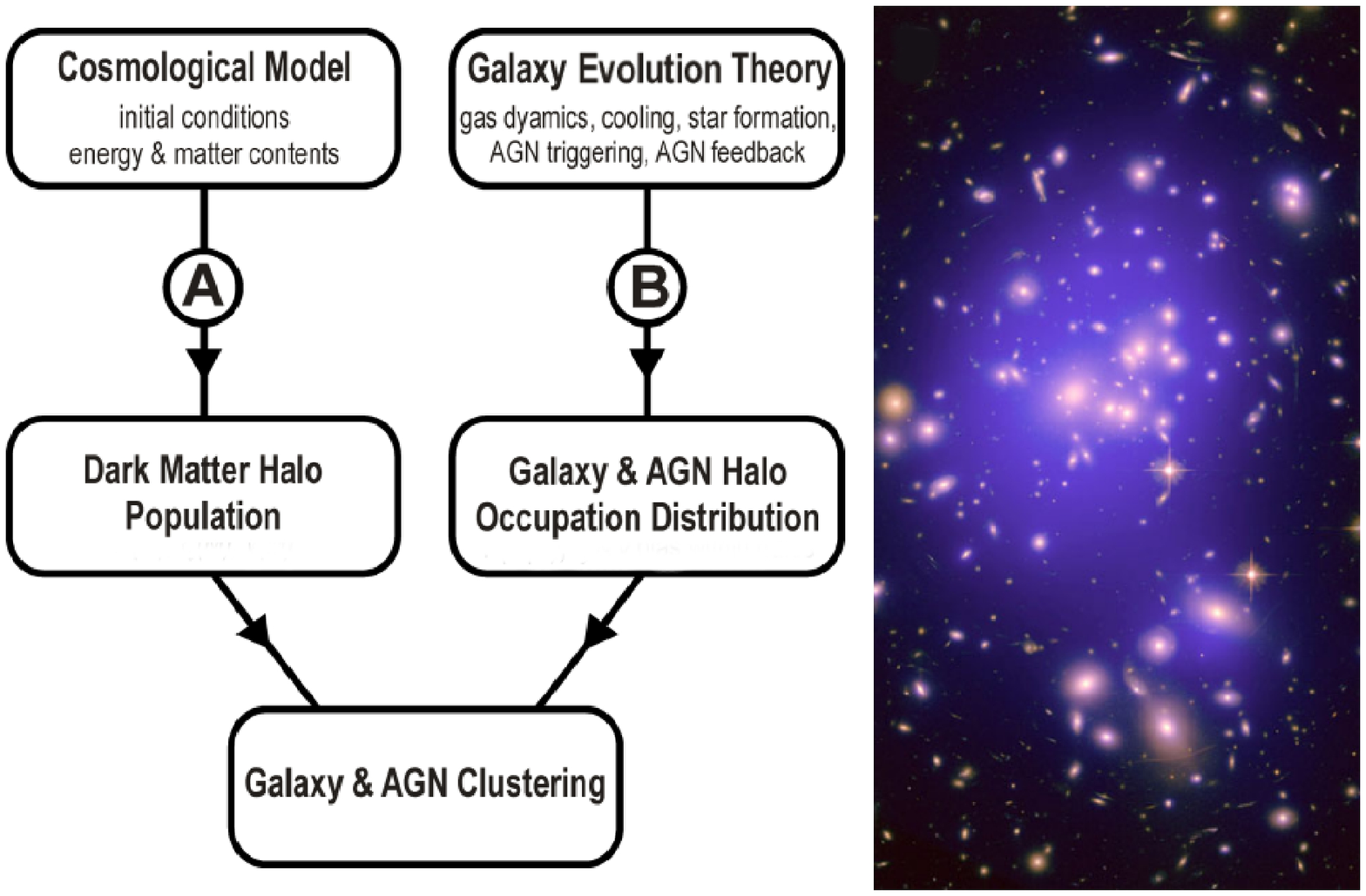}}}
\caption{Current conceptual model of the physical processes involved in 
large-scale galaxy and AGN clustering. {\it Left:} The two branches (A and B) 
in the diagram show the primary causes of clustering: (A) the properties of the 
dark matter halo population, which are 
based on the cosmological model, and (B) the physics 
of complex processes in galaxy formation and evolution, which lead to a 
distinct baryonic population within collapsed dark matter halos. Figure adapted 
from Weinberg 2002.
{\it Right:} Illustration of the spatial distribution of galaxies within a dark 
matter halo. The picture maps the galaxy cluster 
Abell 1689, where an optical image showing 
the galaxy cluster members is superimposed with 
the distribution of dark matter shown in purple. 
Credit: NASA, ESA, E. Jullo, P. Natarajan, and J-P. Kneib.}
\label{fig1}
\end{myfigure}

Inside DMHs, or within halos inside another DMH, called sub-halos, the baryonic
gas will 
radiatively cool. If the gas reservoir is large enough, star and galaxy 
formation will be initiated. The gas can also be accreted onto the 
SMBH in the center of the galaxy. 
On scales comparable to the size of the galaxy, the AGN might heat and/or eject the 
surrounding gas, preventing star formation, and eventually removing the gas supply of 
the AGN itself. All the galaxy evolution processes described here determine how
galaxies 
and AGNs are distributed within DMHs (Fig.~\ref{fig1}, left panel, B-branch). 
This distribution of AGN and galaxies within DMHs (Fig.~\ref{fig1}, right panel) 
is described by the halo occupation distribution (HOD; Peacock \& Smith~2000). In
addition to the 
spatial distribution of AGN/galaxies in DMHs, the HOD describes the probability 
distributions of the number of AGN/galaxies per DMH of a certain mass and the
velocity distribution of AGN/galaxies within a DMH.

The interplay between cosmology and galaxy evolution causes the observed large-scale 
clustering of galaxies and AGNs. The goal of AGN and galaxy clustering measurements 
is to reverse the causal arrows in the Fig.~\ref{fig1} (left panel), working backwards 
from the data to the galaxy \& AGN halo occupation distribution and DMH population 
properties, in order to finally draw conclusions about galaxy and 
AGN physics, as well as to constrain fundamental cosmological parameters.

\section{Clustering Measurements}

The most common statistical estimator for large-scale clustering is the two-point
correlation function (2PCF; Peebles~1980) $\xi(r)$. 
This quantity measures the spatial clustering of a class of object 
{\it in excess} of a Poisson distribution. In practice,
$\xi(r)$ is 
obtained by counting pairs of objects with a given separation and comparing them to
the 
number of pairs in a random sample with the same separation. Different correlation
estimators
are described in the literature (e.g., Davis \& Peebles~1983; Landy \& Szalay~1993).

The large-scale clustering of a given class of object can be quantified by 
computing the angular (2D) correlation function, which is the projection 
onto the plane of the sky, or with the spatial (3D) correlation function,  
which requires redshift information for each object.
Obtaining spectra to measure the 3D correlation function is observationally expensive, which is the main reason why some studies have 
had to rely on angular correlation functions.
However, 3D correlation function measurements are by far 
preferable, since the deprojection (Limber~1954)
of the angular correlation function introduces large systematic uncertainties. Despite 
these large caveats and the already moderately low uncertainties of current 3D
correlation 
measurements, the use of angular correlation function might still be justified when
exploring 
a new parameter space. However, the next generation multi-object spectrographs (e.g.,
4MOST (de Jong et~al.~2012), 
BigBOSS (Schlegel et~al.~2012), and WEAVE (Dalton et~al.~2012), will make it far easier 
to simultaneously 
obtain thousands of spectra over wide fields. Hence, measurements of the 3D 
correlation function will soon become ubiquitous.

As one measures line-of-sight distances for 3D correlation functions from 
redshifts, 
measurements of $\xi(r)$ are affected by redshift-space distortions due to peculiar 
velocities of the objects within DMHs. To remove this effect, $\xi(r)$  is 
commonly extracted by counting pairs on a 2D grid 
of separations where $r_{p}$ is perpendicular to the line of sight and 
$\pi$ is along the line of sight. 
Then, integrating along the $\pi$-direction leads to the projected 
correlation function, $w_p(r_p)$, which is free of redshift distortions.  
The 3D correlation function $\xi(r)$ can be recovered from the projected 
correlation function (Davis \& Peebles~1983). 

The resulting signal can be approximated by a power law where the 
largest clustering strength is found at small scales. At large separations of  
$>$50 Mpc $h^{-1}$ the distribution of objects in the universe becomes nearly 
indistinguishable from a randomly-distributed sample. Only on comoving scales
of $\sim$100 Mpc $h^{-1}$ can a weak positive signal be detected (e.g.,
Eisenstein et~al.~2005; Cole et~al.~2005)
which is caused by baryonic acoustic oscillations (BAO) in the early universe.
  
The spatial clustering of observable objects does not precisely mirror the
clustering 
of matter in the universe. In general, the large-scale density distribution 
of an object class is a function 
of the underlying dark matter density. This relation of how an object class 
traces the 
underlying dark matter density is quantified using the linear bias
parameter $b$. 
This 
contrast enhancement factor is the ratio of the mean overdensity of
the observable object class, the so-called tracer set, to the mean overdensity of
the dark matter field, defined as 
$
 b= (\delta \rho/\langle \rho \rangle)_{\rm tracer} / (\delta \rho/\langle \rho \rangle)_{\rm DM},
$
where $\delta \rho = \rho - \langle \rho \rangle$, $\rho$ is the local mass density, 
and $\langle \rho \rangle$ is the mean mass density on that scale. 
In terms of the correlation function, the bias parameter is defined as the 
square root of the 2PCF ratio of the tracer set to the dark matter 
field: $b = \sqrt{\xi_{\rm tracer} / \xi_{\rm DM}}$. 
Rare objects which form only  
in the highest density peaks of the mass distribution have a large bias parameter and 
consequently a large clustering strength. 

Theoretical studies of DMHs (e.g., Mo \& White~1996; Sheth et~al.~2001) have established 
a solid understanding of the bias parameter of DMHs with respect to various 
parameters. Comparing the bias parameter of an object class with that of 
DMHs in a certain mass range at the same cosmological epoch 
allows one to determine the DMH mass which hosts the object class of interest. 
A halo may contain substructures, but the DMH mass inferred from the linear bias
parameter refers to the single, largest (parent) halo in the context of HOD models.

\subsection{Why are we interested in AGN clustering?}

AGN clustering measurements explore different physics on different scales. 
At scales up to the typical size of a DMH ($\sim 1-2$ Mpc), clustering 
measurements are sensitive to the physics of galaxy/AGN formation and evolution. 
Constraints on the galaxy/AGN merger rate and the radial distribution of these 
objects within DMHs can be derived. On scales larger than the size of DMHs, the
large-scale 
clustering is sensitive to the underlying DM density field, which essentially 
depends only on cosmological parameters. Consequently, with only one 
measurement both galaxy/AGN co-evolution and cosmology can be studied.  
 
Future high precision AGN clustering measurements have the potential to accurately
establish missing fundamental parameters that 
describe when AGN activity and feedback occur as a function of luminosity 
and redshift.  Since they will precisely determine how DMHs 
are populated by AGN host galaxies, these measurements will also improve our 
theoretical understanding of galaxy/AGN formation and evolution by enabling comparisons 
to galaxy measurements and cosmological simulations. 
Here, we elaborate on some (though not all) of the critical observational
constraints which are provided by AGN clustering measurements: 
\begin{itemize}
 \item{{\it AGN host galaxy}  -- 
AGN clustering measurements determine the host 
galaxy type in a statistical sense for the entire AGN sample, regardless of the AGN's
luminosity. 
Comparing the observed AGN clustering to very accurate galaxy clustering
measurements, which depend
on different galaxy subclasses (morphological, spectral type, luminosity), constrains
the AGN 
host galaxy type.}
\vspace*{-0.2cm}
 \item{{\it External (mergers) vs. internal triggering}  -- 
Different theoretical models (e.g., Fry~1996; Sheth et~al.~2001; Shen~2009)
of how AGNs are triggered predict very different large-scale clustering properties
with AGN parameters such as luminosity and redshift. Moderately precise 
AGN clustering measurements allow us to distinguish between these different models 
(Allevato et~al.~2011).
Furthermore, the validity of different models can be tested for different
luminosities and 
cosmological epochs.}
\vspace*{-0.2cm}
 \item{{\it Fundamental galaxy/AGN physics} -- 
AGN large-scale clustering dependences with various AGN properties could potentially be key in providing independent constraints on galaxy/AGN physics. 
Comparing the observed AGN clustering properties with 
results from simulations with different inputs for galaxy/AGN physics  
could identify the physics that links 
the evolution of AGNs and galaxies.}
\vspace*{-0.2cm}
 \item{{\it AGN Lifetimes} -- AGN clustering measurements allow us to estimate 
the AGN lifetime at different cosmological epochs (Martini \& Weinberg~2001). 
The underlying idea is that rare, massive DMHs are highly biased tracers of the 
underlying mass distribution, while more common objects are less strongly biased 
(Kaiser~1984). Therefore, if AGNs are heavily biased
they must be in rare, massive DMHs. The ratio of the AGN number density to
the host halo number density is a measure of the ``duty cycle'', i.e., the fraction
of the time that the object spends in the AGN phase.}
\vspace*{-0.11cm}
 \item{{\it Cosmological parameters}  --
As AGN clustering measurements extend to much higher redshifts than galaxy clustering measuring, they 
can be used to derive constraints on cosmological parameters back to the time 
of the formation of the first AGNs. Currently, the detection of the BAO imprint  on clustering measurements at different cosmological epochs 
is of great interest to constrain the equation of state of dark energy. 
AGN large-scale clustering measurements with very large AGN samples 
can detect the BAO signal in a redshift range that is not accessible with galaxy
clustering 
measurements.}
\end{itemize}

\section{AGN Clustering Measurements: Past and Present}
Until the 1980s, studies had to primarily rely on small, optically-selected, 
very luminous AGN samples for clustering measurements. Then 
the main question was whether AGNs are randomly distributed in the universe 
(e.g., Bolton et~al.~1976; Setti \& Woltjer~1977).  
The extremely small samples sizes did not allow clustering measurements
at scales below $\sim$50 Mpc, where a significant deviation from a random 
distribution is present. Thanks to the launch of major 
X-ray missions in the 1980s and 1990s such as Einstein (Giacconi et~al.~1979) 
and ROSAT (Truemper~1993), much larger 
AGN samples enabled successful detections of the AGN large-scale clustering signal.
A detailed review on the history of X-ray AGN clustering measurements is given 
in Cappelluti et~al.~(2012).

Although AGN clustering measurements are far from being as precise as galaxy
clustering measurements, some general findings have emerged in recent years. 
Interestingly, 
over all of studied cosmic time ($z\sim 0-3$) 
broad-line AGNs occupy DMH masses of 
log $(M_{\rm DMH}/[h^{-1} M_{\odot}])\sim 12.0-13.5$ and therefore cluster like
groups of galaxies. More detailed information about the current
picture of broad-line AGN clustering is presented in Section~6.6 of 
Krumpe et al.~(2012).

Some puzzling questions remain. For example, at $z<0.5$ a weak X-ray 
luminosity dependence on the clustering strength is found (in that 
luminous X-ray AGNs 
cluster more strongly than their low luminosity counterparts, e.g., 
Krumpe et~al.~2010; Cappelluti et~al.~2010; Shen et~al.~2012). 
However, at high redshift it seems that high luminosity,  
optically-selected AGNs cluster less strongly 
than moderately-luminous X-ray selected AGNs.
Whether this finding 
is due to differences in the AGN populations, an intrinsic 
luminosity dependence to the clustering amplitude, or an observational 
bias is yet not understood. 

We note that different studies have used different relations to translate 
the measured linear bias parameter to DMH mass, as well as different $\sigma_8$
values. 
Therefore, instead of blindly comparing the derived DMH mass, re-calculating the
masses based on 
the same linear bias to DMH mass relation and the same $\sigma_8$ is essential when
comparing measurements in the literature.

\subsection{Recent Developments}
In the last few years several new approaches have been used to improve 
the precision of AGN clustering measurements or their interpretation. 
We briefly summarize these developments below.
 
\vspace*{0.1cm}
\hspace*{-0.5cm}{\it Cross-correlation measurements: }\\
Auto-correlation function (ACF) measurements of broad-line AGNs often 
have large uncertainties due to the low number of 
objects. Especially at low redshifts, large galaxy samples
 with spectroscopic redshifts are 
frequently available. In such cases, the statistical uncertainties of AGN 
clustering measurements can be reduced significantly by computing the 
cross-correlation function (CCF). The CCF measures
the clustering of objects between two different object classes 
(e.g., broad-line AGNs and galaxies), while the ACF measures the spatial
clustering of 
objects in the same sample (e.g., galaxies or AGNs). CCFs have been used before to 
study the dependence of the AGN clustering signal with different AGN 
parameters. However, 
these values could not be compared to other studies as the CCFs also depend 
on the galaxy populations used and their clustering strength. Only 
recently has an alternative approach (Coil et~al.~2009) allowed the comparison 
of the results from different studies by inferring the AGN ACF from the 
measured AGN CCF and ACF of the galaxy tracer set. The basic idea of this 
method, which is now frequently used (e.g., 
Krumpe et~al.~2010, 2012; Mountrichas \& Georgakakis~2012; Shen et~al.~2012), 
is that both populations trace the same underlying DM density field. 

\vspace*{0.15cm}
\hspace*{-0.5cm}{\it Photometric redshift samples: }\\
Large galaxy tracer sets with spectroscopic redshifts 
are not available at all redshifts. Some studies therefore rely on 
photometric redshifts. The impact of the large uncertainties and 
catastrophic outliers when using photometric redshifts 
is commonly not considered but it is 
essential. The use of the full probability density function (PDF) of the
photometric 
redshift fit, instead of a single value for the photometric redshift, has been 
used in some studies (e.g., Mountrichas et~al.~2013). 
Here photometric galaxies samples 
are used as tracer sets to derive the CCF. Each object is given a 
weight for the probability that it is actually located at a certain redshift based on 
the fit to the photometric data.

\begin{myfigure}
\centerline{\resizebox{80mm}{!}{\includegraphics{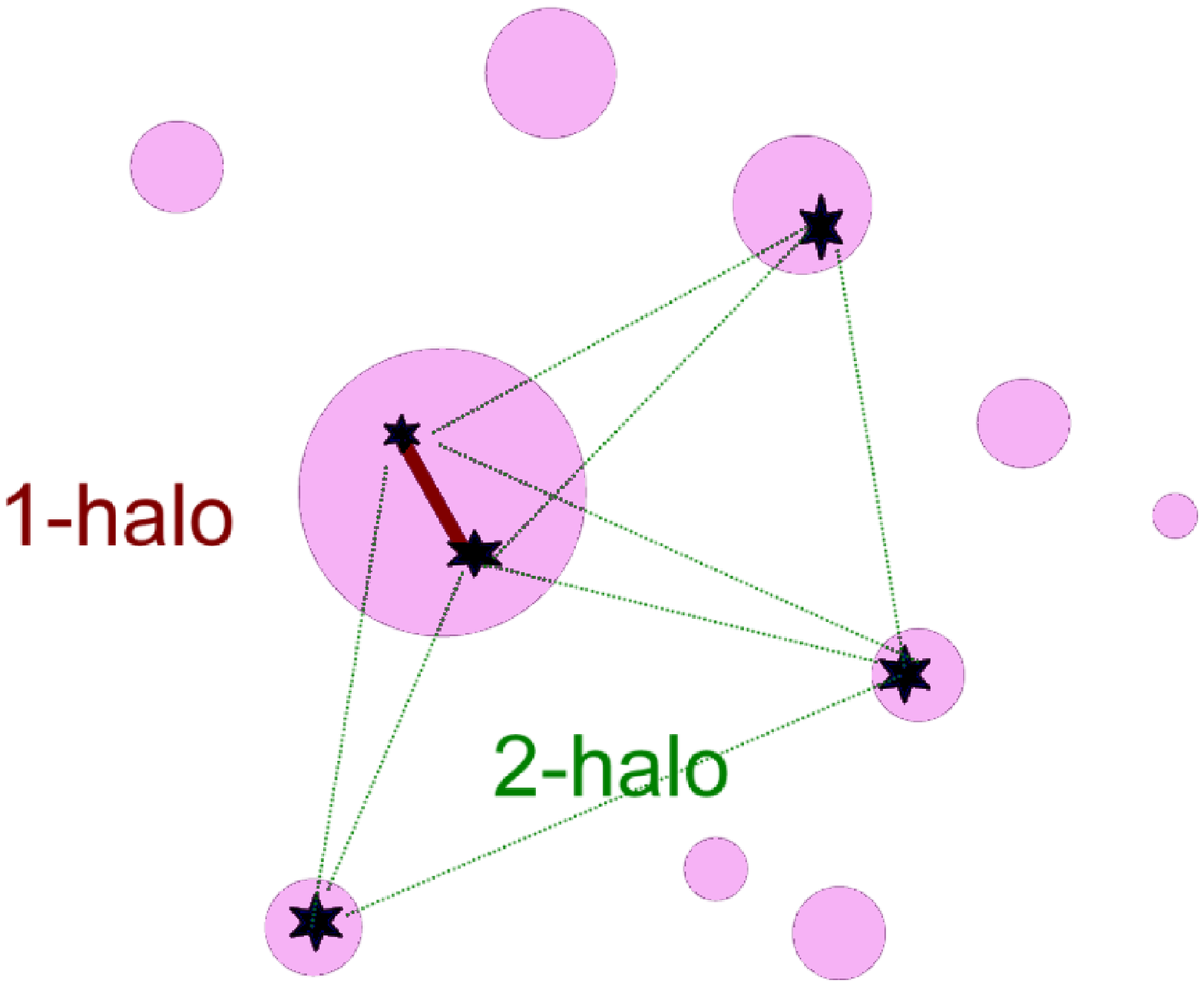}}}
\caption{ In the conceptual model of the HOD approach, there are
two contributions to the pairs that 
account for the measured correlation function. 
Pairs of objects (black stars) can either be located within the same 
DMH (pink filled circles), such that their measured separation contributes to 
the 1-halo term (red solid line in the large DMH), or can reside 
in different DMHs, such that 
their separations (green dotted line) contribute to the 
2-halo term.}
\label{hod}
\end{myfigure}

\hspace*{-0.5cm}{\it AGN Halo Occupation Distribution Modeling: }\\
Instead of deriving only mean DMH masses from the linear bias parameter, HOD
modeling of the correlation function allows the determination of the full 
distribution of AGN as a function of DMH mass. The derived distribution 
also connects observations and simulations as it provides recipes for how 
to populate DMHs with observable objects.

In the HOD approach, the measured 2PCF is modeled as the sum of contributions 
from pairs within individual DMHs (Fig.~\ref{hod}; 1-halo term) and in 
different DMHs 
(2-halo term). The superposition of both components describes the shape of 
the observed 2PCF better than a simple power law.
In the HOD description, a DMH can be populated by one central AGN/galaxy and 
by additional objects in the same DMH, so-called satellite AGN/galaxies.
Applying the HOD approach to the 2PCF allows one to determine, e.g., the minimum DMH 
needed to host the object class of interest, the fraction of objects in satellites, 
and the number of satellites as a function of DMH mass. 
Instead of using the derived AGN ACF from CCF measurements, 
Miyaji et~al.~(2011) utilize the HOD model directly on high precision AGN/galaxy CCF
and achieve additional constraints on the AGN/galaxy co-evolution and AGN physics.

\hspace*{-0.5cm}{\it Theoretical predictions: }\\
Only recently have several different theoretical models been published which 
try to explain 
the observed AGN clustering with different physical approaches 
(e.g, Fanidakis et~al.~2013a; H\"utsi et~al.~2013). 
The key to observationally distinguishing between these models are their different 
predictions for the clustering dependences of different AGN parameters. 
In addition to theoretical models of the observed clustering, other very recently 
developed models predict the halo occupation distribution of AGN at different 
redshifts, e.g., Chatterjee et~al.~(2012). 
The major challenge presented by all of these models is the urgent need for 
observational constraints 
with higher precision than can be provided with current AGN samples.
In the future, progress in AGN physics and AGN/galaxy evolution 
will be achieved through a close interaction between state-of-the-art cosmological 
simulations and observational constraints from high precision clustering measurements.
Simulations which incorporate different physical processes will 
lead to different predictions of the AGN and galaxy large-scale clustering trends and 
their halo occupation distributions. 
Observational studies will then identify the correct model and consequently 
the actual underlying physical processes.

\section{The future of AGN clustering measurements}
AGN clustering measurements from several upcoming projects will significantly  
extend our knowledge of the growth of 
cosmic structure and will also provide a promising avenue towards new 
discoveries in the fields of galaxy/AGN co-evolution, AGN triggering, 
and cosmology. For example, eROSITA (Predehl et~al.~2010); launch
2014/2015) 
will perform several all-sky X-ray surveys. 
After four years the combined survey is expected to 
contain approximately three million AGNs.
HETDEX (Hill et~al.~2008) will use an array of integral-field spectrographs to 
provide a total sample of $\sim$20,000 AGNs without any pre-selection over an 
area of 
$\sim$ 420 deg$^2$. The SDSSIV/eBOSS and BigBOSS builds
upon the SDSS-III/BOSS project
and will use a fiber-fed spectrograph. Over an area of 14,000 deg$^2$, 
it will observe roughly one million QSOs at $1.8 < z <3.5$. In addition to these projects, 
there will be other major enterprises such as LSST (LSS collaboration~2009) and 
Pan-STARRS (Kaiser et~al.~2002) which will detect several million AGNs (though
these surveys currently lack dedicated spectroscopic follow-up programs). 

In the following we will focus on eROSITA, as this mission will compile the largest 
AGN sample ever observed. Figure~\ref{fig2} shows that
eROSITA AGN detections will outnumber at $z>0.4$ current galaxy samples 
with spectroscopic redshifts.
Using a large number of AGNs that continuously cover the redshift space, 
will allow us (in contrast to galaxy samples) to measure the distribution of
matter with high precision in the last $\sim$11 Gyr of cosmic time.
To fully exploit the eROSITA potential for AGN clustering 
measurements, a massive spectroscopic follow-up program is needed. Several 
spectroscopic multi-object programs and instruments are currently planned or are in 
an early construction phase (e.g., SDSS IV/SPIDERS and 4MOST).

\begin{myfigure}
\centerline{\resizebox{88mm}{!}{\includegraphics{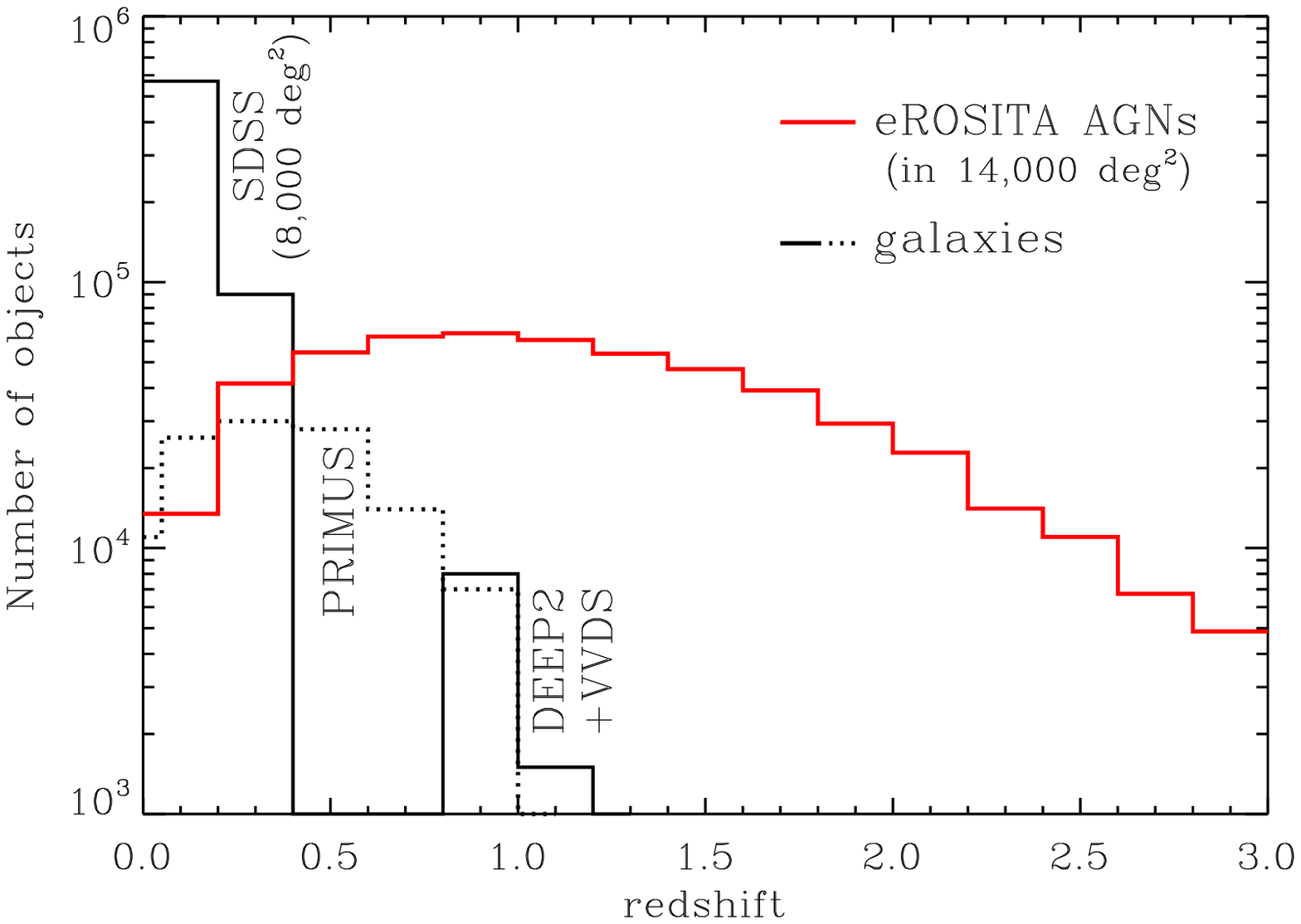}}}
\caption{Number of expected eROSITA AGNs (red) and currently available 
galaxies with spectroscopic redshifts
(black solid line at $z<0.4$ -- SDSS data release 7; black dotted line -- PRIMUS
(Coil et~al.~2011); black solid line at $z\sim1$ -- DEEP2 (Davis et~al.~2003) 
and VVDS (Le F\`evre et~al.~2005)). 
Instead of the full sky area, we consider only the expected number of eROSITA 
AGNs with spectroscopic redshifts from 4MOST over 14,000 deg$^2$.}
\label{fig2}
\end{myfigure}

eROSITA AGN clustering measurements at $z \sim 0.8-1$ will even allow for the 
detection of the BAO signal. The feasibility of such a measurement can be 
estimated using the BAO detection found with $\sim$46,000 SDSS LRGs 
($\langle$$z$$\rangle=0.35$) over 3,816 square 
degrees of sky (0.72 $h^{-3}$ Gpc$^3$) as a standard for comparison
(Eisenstein et~al.~2005).
The observed AGN X-ray luminosity function (Gilli et~al.~2007) and the  
eROSITA sensitivity determine the number density of eROSITA AGNs. 
In the above redshift range, the eROSITA AGN area density will be 
comparable to that of SDSS LRGs at lower redshifts. Therefore, the comoving volume number 
density of eROSITA AGNs will be five times lower than that of SDSS LRGs. 
Since eROSITA will conduct an all-sky survey, the increased sky area will counterbalance 
the lower volume density. Given the signal-to-noise ratio (S/N) of the 
BAO detection of Eisenstein et al. (2005) and an assumed spectroscopic area of 
14,000 deg$^2$, we expect a $\sim$3$\sigma$ BAO detection using eROSITA AGNs only 
in the redshift range of $z \sim 0.8-1$. This is consistent with 
Kolodzig et~al.~(2013), who use a different approach based on the angular power 
spectrum for estimating the significance of a BAO detection with eROSITA AGN.

With the much larger AGN datasets that will exist in the future, 
the statistical uncertainties in clustering measurements
will be significantly decreased. Systematic uncertainties will then be the 
dominant source of uncertainty. The impact and level of different systematic
uncertainties
can only be carefully explored and quantified through simulations. 
Thus far, there has not been a need for such studies because 
the AGN samples to date are i) drawn from surveys that (with exceptions) 
cover a rather moderate sky area and are therefore likely to suffer from the problem 
of cosmic 
variance\footnote{Surveys with a small sky area will not sample a representative part of the
universe due to cosmic variance, i.e., if a survey incidentally aims at an underdensity/overdensity in the 
universe a lower/higher clustering amplitude will be measured than when aiming 
at a representative part of the universe.}
and/or ii) comprised of up to several thousand 
objects and are consequently Poisson noise dominated. 
Both limitations will be removed in future AGN clustering measurements with 
the upcoming extensive AGN samples covering extremely large sky areas. 
However, to derive reliable constraints on AGN physics and cosmology, 
as well as to avoid any possible misinterpretations of future unprecedented high precision AGN clustering
measurements, we have to fully understand and be able to correctly model the impact of the
systematic uncertainties. 
Only then can we maximize the scientific return of future AGN clustering
measurements and have a major impact in the field of cosmology and galaxy/AGN evolution.

\thanks
MK received funding from the European Community's Seventh Framework Programme 
(/FP7/2007-2013/) under grant agreement No 229517. TM is supported by 
UNAM/PAPIIT IN104113 and CONACyT 179662. ALC acknowledges support from NSF 
CAREER award AST-1055081.

\end{multicols}

\vspace*{0.0cm}
\begin{thebibliography}{99}
\vspace*{-0.05cm}
\bibitem{abazajian_2009} Abazajian et al. 2009, ApJS, 182, 543.\vspace*{-0.2cm}
\bibitem{allevato_2011} Allevato et al. 2011, ApJ, 736, 99.\vspace*{-0.2cm}
\bibitem{bolton_1976} Bolton et al. 1976, ApJ, 210, L1.\vspace*{-0.2cm}
\bibitem{cappelluti_2010} Cappelluti et al. 2010, ApJ, 716, L209.\vspace*{-0.2cm}
\bibitem{cappelluti_2012} Cappelluti et al. 2012, AdAst, 2012, 25. \vspace*{-0.2cm}
\bibitem{chatterjee_2012} Chatterjee et al. 2012, MNRAS, 419, 2657. \vspace*{-0.2cm}
\bibitem{coil_2009} Coil et al. 2009, ApJ, 701, 1484.\vspace*{-0.2cm}
\bibitem{coil_2011} Coil et al. 2011, ApJ, 741, 8.\vspace*{-0.2cm}
\bibitem{cole_2005} Cole et al. 2005, MNRAS, 362, 505.\vspace*{-0.2cm}
\bibitem{colless_2001} Colless et al. 2001, MNRAS, 328, 1039.\vspace*{-0.2cm}
\bibitem{dalton_2012} Dalton et al. 2012, SPIE, 8446.\vspace*{-0.2cm}
\bibitem{davis_1983} Davis \& Peebles 1983, ApJ, 267, 465.\vspace*{-0.2cm}
\bibitem{davis_2003} Davis et al. 2003, SPIE, 4834, 161.\vspace*{-0.2cm}
\bibitem{deJong_2012} de Jong et al. 2012, SPIE, 8446.\vspace*{-0.2cm}
\bibitem{eisenstein_2005} Eisenstein et al. 2005, ApJ, 633, 560.\vspace*{-0.2cm}
\bibitem{fanidakis_2013} Fanidakis et al. 2013, arXiv:1305.2200.\vspace*{-0.2cm}
\bibitem{ferrarese_2000} Ferrarese \& Merritt 2000, ApJ, 539, L9.\vspace*{-0.2cm} 
\bibitem{fry_1996} Fry 1996, ApJ, 461L, 65.\vspace*{-0.2cm}
\bibitem{gebhardt_2000} Gebhardt et al. 2000, ApJ, 539, 13.\vspace*{-0.2cm}
\bibitem{giacconi_1979} Giacconi et al. 1979, ApJ, 230, 540.\vspace*{-0.2cm}
\bibitem{gilli_2007} Gilli et al. 2007, A\&A, 463, 79.\vspace*{-0.2cm}
\bibitem{hasinger_2005} Hasinger et al. 2005, A\&A, 441, 417.\vspace*{-0.2cm}
\bibitem{hill_2008} Hill et al. 2008, ASPC, 399, 115.\vspace*{-0.2cm}
\bibitem{hopkins_2006} Hopkins \& Beacom 2006, ApJ, 651, 142.\vspace*{-0.2cm}
\bibitem{huetsi_2013} H\"utsi et al. 2013, arXiv:1304.3717.\vspace*{-0.2cm}
\bibitem{jahnke_2011} Jahnke \& Macci\'o 2011, ApJ, 734, 92.\vspace*{-0.2cm}
\bibitem{kaiser_1984} Kaiser 1984, ApJ, 284, L9.\vspace*{-0.2cm}
\bibitem{kaiser_2002} Kaiser et al. 2002, SPIE, 4836, 154.\vspace*{-0.2cm}
\bibitem{kolodzig_2013} Kolodzig et al. 2013, arXiv:1305.0819\vspace*{-0.2cm}
\bibitem{krumpe_2010} Krumpe et al. 2010, ApJ, 713, 558.  \vspace*{-0.2cm}
\bibitem{krumpe_2012} Krumpe et al. 2012, ApJ, 746, 1.  \vspace*{-0.2cm}
\bibitem{landy_1993} Landy \& Szalay 1993, ApJ, 412, 64.\vspace*{-0.2cm}
\bibitem{larsen_2011} Larsen et al. 2011, ApJS, 192, 16.\vspace*{-0.2cm}
\bibitem{lefevre_2005} Le F\`evre et al. 2005, A\&A, 439, 845.\vspace*{-0.2cm}
\bibitem{limber_1954} Limber 1954, ApJ, 119, 655.\vspace*{-0.2cm}
\bibitem{lsst_2009} LSST Collaboration 2009, arXiv0912.0201.\vspace*{-0.2cm}
\bibitem{magorrian_1998} Magorrian et al. 1998, AJ, 115, 2285.\vspace*{-0.2cm}
\bibitem{martini_2001} Martini \& Weinberg 2001, ApJ, 547, 12.\vspace*{-0.2cm}
\bibitem{miyaji_2011} Miyaji et al. 2011, ApJ , 726, 83.\vspace*{-0.2cm}
\bibitem{mo_1996} Mo \& White 1996, MNRAS, 282, 347.\vspace*{-0.2cm}
\bibitem{mountrichas_2012} Mountrichas \& Georgakakis 2012, MNRAS, 420, 514.\vspace*{-0.2cm}
\bibitem{mountrichas_2013} Mountrichas et al. 2013, MNRAS, 430, 661.\vspace*{-0.2cm}
\bibitem{peacock_2000} Peacock \& Smith 2000, MNRAS, 318, 1144.\vspace*{-0.2cm}
\bibitem{peebles_1980} Peebles 1980, Princeton University Press.\vspace*{-0.2cm}
\bibitem{predehl_2010} Predehl et al. 2010, SPIE, 7732, 23.\vspace*{-0.2cm}
\bibitem{schlegel_2011} Schlegel et al. 2011, arXiv1106.1706.\vspace*{-0.2cm}
\bibitem{setti_1977} Setti \& Woltjer 1977, ApJ, 218, L33.\vspace*{-0.2cm}
\bibitem{shen_2009} Shen 2009, ApJ, 704, 89.\vspace*{-0.2cm}
\bibitem{shen_2012} Shen et al. 2012, arXiv1212.4526S.\vspace*{-0.2cm}
\bibitem{sheth_2001} Sheth et al. 2001, MNRAS, 323, 1.\vspace*{-0.2cm}
\bibitem{smoot_1992} Smoot et al. 1992, ApJ, 396, 1.\vspace*{-0.2cm}
\bibitem{truemper_1993} Truemper 1993, Sci., 260, 1769.\vspace*{-0.2cm}
\bibitem{weinberg_2002} Weinberg 2002, ASPC, 283, 3.\vspace*{-0.2cm}
\end{thebibliography}
\end{document}